\documentclass[12pt]{article}
\usepackage{amsfonts}
\usepackage{amsmath}
\usepackage{amssymb}
\usepackage{graphicx}
\usepackage{pst-node}
\usepackage{tikz-cd} 
\usepackage{color}
\usepackage[margin=3cm]{geometry}

\def \be {\begin{equation}}
\def \ee {\end{equation}}
\def \bea {\begin{eqnarray}}
\def \eea {\end{eqnarray}}
\def \nn {\nonumber}

\def \rr {\raise.35ex\hbox{\small $\prime$}\kern-.17em{\mbox{\large $\imath$}}}

\def \dels {\partial\kern-.6em /\kern.1em}
\def \As {{A\kern-.5em / \kern.5em}}
\def \Ds {D\kern-.7em / \kern.5em}

\def \ks {k\kern-.5em /}
\def \ls {l\kern-.5em /}

\newcommand{\hide}[1]{}

\reversemarginpar
\begin{document}

\begin{titlepage}

\begin{center}

\hfill
\vskip .2in

\textbf{\LARGE
A Duality Web in Condensed Matter Systems
\vskip.3cm
}

\vskip .5in
{\large
Chen-Te Ma \footnote{e-mail address: yefgst@gmail.com} 
\\
\vskip 1mm
}
{\sl
Department of Physics and Center for Theoretical Sciences, \\
National Taiwan University,\\ 
Taipei 10617, Taiwan, R.O.C.
}\\
\vskip 1mm
\vspace{40pt}
\end{center}
\begin{abstract}
We study various dualities in condensed matter systems. The dualities in three dimensions could be derived from a conjecture of a duality between a Dirac fermion theory and an interacting scalar field theory at the Wilson-Fisher fixed point and zero temperature in three dimensions. We show that the dualities are not affected by the non-trivial holonomy, use a mean-field method to study, and discuss the dualities at a finite temperature. Finally, we combine a bulk theory, which is an Abelian $p$-form theory with a theta term in $2p+2$ dimensions, and a boundary theory, which is a $2p+1$ dimensional theory, to discuss constraints and difficulties of a duality web in $2p+1$ dimensions. 
\end{abstract}

\end{titlepage}

\section{Introduction}
\label{1}
Duality could exhibit non-trivial equivalence between two different theories. A generic study of a duality web appears in string and supergravity theory \cite{Giveon:1994fu}. String and supergravity theory shows a beautiful duality web to unify different fundamental theories through T-duality and S-duality \cite{Witten:1995gf}. S-duality could exchange a weak coupling constant and a strong coupling constant to demonstrate equivalence between a weakly coupled effective field theory and a strongly coupled effective field theory. Hence, S-duality could show non-perturbative physics through a perturbative study. The electric-magnetic duality in an Abelian $p$-form gauge theory in $2p+2$ dimensions is studied in \cite{Witten:2003ya, Ho:2015mfa}.

A duality web in condensed matter systems is founded in \cite{Seiberg:2016gmd} for a generic background at an infrared (IR) limit and zero temperature  in three dimensions by considering a $Spin_c$ manifold. The motivation of introducing a $Spin_c$ manifold comes from an inconsistent dualization of a conjecture \cite{Seiberg:2016gmd}, which is equivalence between a Dirac fermion and an interacting scalar field theory, in a $Spin$ manifold. Because we expect that the conjecture should be correct for a generic background, the dualization of the conjecture should not show inequivalent dependence of a choice of a spin structure \cite{Seiberg:2016gmd}. Therefore, a generalization from a $Spin$ manifold to a $Spin_c$ manifold is useful for using a larger structure to study three dimensional  condensed matter systems. The related study of the duality web for a flat background is in \cite{Karch:2016sxi}, the duality web at a finite temperature is in \cite{Ma:2016yas}, a duality between a four dimensional topological insulator and a topological superconductor is in \cite{Murugan:2016zal}, dualities for an $SO$ gauge group are in \cite{Metlitski:2016dht, Aharony:2016jvv} and dualities for a $USp$ gauge group are in \cite{Aharony:2016jvv}, a lattice study in the particle-vortex duality is in \cite{Peskin:1977kp}, and a derivation of dualities by promoting background fields to dynamical fields is in \cite{Karch:2016aux}. Other useful studies of a mirror symmetry for a supersymmetric defect is in \cite{Hook:2013yda}, a half-filled Landau level is in \cite{Kachru:2015rma}, a bosonization is in \cite{Kachru:2016rui} and non-supersymmetric dualities are in \cite{Kachru:2016aon}.

The effective field theory in the conjecture is an Abelian Chern-Simons theory, which is a topological quantum field theory, at an IR limit in three dimensions. The Abelian Chern-Simons theory with a one-half coefficient could be generated from a one-loop effective potential of a massless Dirac fermion with a regularization \cite{Atiyah:1975jf} in three dimensions \cite{Niemi:1983rq}. The massless Dirac fermion does not have a parity violation in three dimensions at classical level, but the parity anomaly appear at  quantum level \cite{Redlich:1983dv}. The parity anomaly with gravitation is also studied in \cite{AlvarezGaume:1984nf} and the parity anomaly in an unorientable manifold is discussed in \cite{Witten:2016cio}. A discussion of a canonical quantization of the Abelian Chern-Simons theory is in \cite{Elitzur:1989nr} and a holomorphic view of three dimensional topological quantum field theory is in \cite{Witten:1988hf}. More useful studies of fermion path integrals are in \cite{Witten:2015aba} and topological phases are in \cite{Witten:2015aba, Seiberg:2016rsg}

The three dimensional duality web also could be understood from modular or $SL(2)$ transformations in a four dimensional Abelian gauge theory with a theta term in a manifold with a boundary \cite{Seiberg:2016gmd} at an IR limit. When we consider the four dimensional theory or the bulk theory and the three dimensional Dirac fermion theory or the boundary theory simultaneously, we could find that the four dimensional dynamical gauge field becomes a background gauge field by taking an IR limit, which leads a weak four dimensional coupling constant. The IR limit also leads a strong three dimensional coupling constant. Three dimensional dynamical gauge potential is still dynamical at the IR limit and the three dimensional gauge potential is also coupled to a scalar field. The scalar field theory is another boundary theory, in which the gauge potential is dynamical. The combination of the bulk theory and the boundary theory could be studied self-consistently. The study also demonstrates a more reliable result. If we only consider three dimensions or a boundary theory at the IR limit, the duality web needs to begin from the conjecture. Thus, all results depend on whether the conjecture is correct. Form the combination of the bulk theory and the boundary theory, we could obtain alternative evidences for the conjecture. The demonstration of the three dimensional particle-vortex duality from the four dimensional Abelian gauge theory with a theta term is in \cite{Metlitski:2015eka}.

In this paper, we discuss various dualities in condensed matter systems. We begin from the conjecture to show other dualities without losing the holonomy. The dualization is to integrate gauge fields out, which is similar to solving $da=db$, where $a$ and $b$ are gauge fields. A general solution of $da=db$ is $a=b+df$ with $d^2f=0$ when $a$ and $b$ satisfy a Dirac quantization condition, which is $\int da=\int db=2\pi$. When we set $f=0$ or zero holonomy due to a gauge symmetry of a physical system, all fields of the physical system should be dynamical. If some fields are just background fields, the holonomy may deform the background fields after we do a dualization. Thus, the three dimensional duality web may be modified by the holonomy. However, we find that a non-trivial holonomy could be absorbed by a scalar field when a boundary term is absent \cite{Ma:2016yas}. Thus, the non-trivial holonomy does not modify the three dimensional duality web when a boundary term is ignored and all gauge fields satisfy a Dirac quantization condition or fields are globally defined.

We also use a mean-field method to study the three dimensional duality web. The study of the mean-field method also begins from the conjecture and use the order parameter $\bar{\psi}\psi$. Our study shows that the order parameter in a Dirac fermion theory could dual to a bosonic mass term in a scalar field theory \cite{Ma:2016yas}. Thus, it is interesting to find an operator correspondence between the fermion theory and the boson theory.

We also propose the inclusion of a finite temperature in the three dimensional duality web. One difficulty of the duality web at a finite temperature is that a gauge invariant effective field theory from that a Dirac fermion needs to do resummation to all orders. The duality web at zero temperature is only guaranteed at the Wilson-Fisher fixed point or an IR limit. In other words, the effective theory to all orders from resummation possibly does not exist a duality web. Hence, our proposal only considers a leading non-trivial order term. Because the form of the effective theory at a finite temperature is the same as the effective theory at zero temperature, except for a coefficient, which breaks a gauge symmetry of the effective theory, a duality web at a finite temperature should work as in the three dimensional duality web at zero temperature.

We also use an Abelian $p$-form gauge theory with a theta term in $2p+2$ dimensions to discuss constraints and difficulties of a duality web in $2p+1$ dimensions at an IR limit. The study is based on a duality web or a boundary theory could be related to a duality of a bulk theory at an IR limit. When we do an $SL(2)$ transformation in a $2p+2$ dimensional manifold with a boundary, we could find a topological term in $2p+1$ dimensions. Thus, the study should give us some constraints to understand construction of a duality web in $2p+1$ dimensions.

We discuss the three dimensional duality web at zero temperature without ignoring the holonomy in Sec.~\ref{2}. The mean-field study of the three dimensional duality web at zero temperature with the order parameter $\bar{\psi}\psi$ is shown in Sec.~\ref{3} and the three dimensional duality web at a finite temperature is exhibited in Sec.~\ref{4}. We also discuss a duality web at zero temperature in other dimensions in Sec.~\ref{5}. Finally, we discuss and conclude in Sec.~\ref{6}. 

\section{A Duality Web in Three Dimensions}
\label{2}
We first show our notation to conveniently express partition functions and the actions and start from a conjecture between a Dirac fermion and an interacting scalar field theory \cite{Seiberg:2016gmd, Karch:2016sxi}. Then we could use the conjecture to derive various dualities in three dimensions at an IR limit. 

\subsection{Notation}
We define the notation of the action of the Abelian Chern-Simons theory with the level one
\bea
S_{CS}[A]=\frac{1}{4\pi}\int d^3x\ \epsilon^{\mu\nu\rho}A_{\mu}\partial_{\nu}A_{\rho},
\eea
the action of the BF theory
\bea
S_{BF}[a ; A]=\frac{1}{2\pi}\int d^3x\ \epsilon^{\mu\nu\rho}a_{\mu}\partial_{\nu}A_{\rho},
\eea
in which we also have
\bea
S_{BF}[a; A]=S_{BF}[A; a]
\eea
up to a boundary term, the action of the scalar field theory
\bea
S_{scalar}[\phi; A]=\int d^3x\ \big(|(\partial_{\mu}+iA_{\mu})\phi|^2-\lambda|\phi|^4\big),
\eea
the action of the massless Dirac fermion theory
\bea
S_{fermion}[\psi; A]=\int d^3x\ i\bar{\psi}\gamma^{\mu}(\partial_{\mu}+iA_{\mu})\psi,
\eea
the partition function of the massless Dirac fermion
\bea
Z_{fermion}[A]=\int D\psi\ \exp\big(iS_{fermion}[\psi ;A]\big),
\eea
the partition function of the scalar field theory
\bea
Z_{scalar}[A]=\int D\phi\ \exp\big(iS_{scalar}[\phi ;A]\big),
\eea
in which we used the metric $\eta_{\mu\nu}=\mbox{diag}(1, -1, -1,\cdots, -1)$ to do contraction.

\subsection{Conjecture of Boson-Fermion Duality}
We start from a conjecture of a boson-fermion duality
\bea
Z_{fermion}[A]\exp\bigg(-\frac{i}{2}S_{CS}[A]\bigg)=\int D\phi Da\ \exp\big(i S_{scalar}[\phi; a]+iS_{CS}[a]+iS_{BF}[a; A]\big)
\nn\\
\eea
in three dimensions. Now we interpret the conjecture. The massless Dirac fermion is not gauge invariant so we need to add the Abelian Chern-Simons theory with the level one-half to restore the gauge symmetry. The Abelian Chern-Simons theory with the level one-half could cancel the parity anomaly to give a consistent theory. 

The equation of motion of $a_0$
\bea
\rho_{scalar}+\frac{f_{12}}{2\pi}=0, \qquad f_{12}\equiv\partial_1a_2-\partial_2 a_1,
\eea
where $\rho_{scalar}$ is the charge density of the complex scalar field, when we turn off the background gauge field $A$. This implies that one unit flux turns on one complex scalar field.

It is also easy to find that a monopole operator of $a$ has $a$ charge one and $A$ charge one and a monopole operator of the complex scalar field $\phi$ has the vanishing $A$ charge and have $a$ charge one-half. Thus, we could use the monopole operators of the complex scalar field and $a$ to construct a gauge invariant operator and identify the gague invariant operator as the fermion field. We could obtain a consistent $A$ charge one for the fermion field.

Because the conjecture does not contain the kinetic term of the Abelian gauge field, it is easy to know that the conjecture must be built at an IR limit or $e^2\rightarrow\infty$, where $e$ is a gauge coupling constant. 

We could rewrite the conjecture:
\bea
&&\int D\psi DA\ \exp\bigg(iS_{fermion}[\psi ; A]-\frac{i}{2}S_{CS}[A]-iS_{BF}[A ;C]\bigg)
\nn\\
&=&\int D\phi Df\ \exp\bigg(i S_{scalar}[\phi; C+df]+iS_{CS}[C+df]\bigg)
\nn\\
&=&e^{iS_{CS}[C]}\int Df\ Z_{scalar}[C+df],
\nn\\
\eea
in which we solve $dA=dC$ to get $A=C+df$ in the boson theory. We also ignored the total derivative term in the second equality. If we consider the equation of the motion of $A_0$ and turn off the background gauge field $C$, we could obtain
\bea
\rho_{fermion}-\frac{1}{2}\frac{F_{12}}{2\pi}=0, \qquad F_{12}\equiv\partial_1 A_2-\partial_2 A_1,
\eea
where $\rho_{fermion}$ is the charge density of the fermion field. Thus, the fermion charge is $1/2$ $Q_{fermion}=1/2$ when  single monopole appears.

We could use the time reversed operation, which flips the sign of the Chern-Simons and $BF$ terms, to obtain
\bea
Z_{fermion}[A]e^{\frac{i}{2}S_{CS}[A]}=\int D\phi Da\ \exp\bigg(iS_{scalar}[\phi ; a]-iS_{CS}[a]-iS_{BF}[a ; A]\bigg)
\eea
and
\bea
&&\int D\psi DA\ \exp\bigg(iS_{fermion}[\psi ; A]+\frac{i}{2}S_{CS}[A]+iS_{BF}[A ;C]\bigg)
\nn\\
&=&e^{-iS_{CS}[C]}\int Df\ Z_{scalar}[C+df].
\nn\\
\eea

\subsection{Boson-Boson Duality}
We derive a boson-boson duality from the conjecture and start from
\bea
&&e^{-iS_{CS}[C]}\int D\psi DA\ \exp\bigg(iS_{fermion}[\psi ; A]-\frac{i}{2}S_{CS}[A]-iS_{BF}[A ;C]\bigg)
\nn\\
&=&\int Df\ Z_{scalar}[C+df].
\nn\\
\eea
Now we add the $BF$ term to get
\bea
&&\int Da\ \exp\bigg(-iS_{CS}[a]+iS_{BF}[a ; A]\bigg)\int D\psi D\tilde{a}
\nn\\
&&\times\exp\bigg(iS_{fermion}[\psi ; \tilde{a}]-\frac{i}{2}S_{CS}[\tilde{a}]-iS_{BF}[\tilde{a} ;a]\bigg)
\nn\\
&=&\int Da\ \exp\bigg(iS_{BF}[a ; A]\bigg)\int Df\ Z_{scalar}[a+df],
\eea
then we could integrate the dynamical gauge field $a$ out in the fermion theory, which is equivalent to getting $a=A-\tilde{a}+dg$, to obtain
\bea
&&\int D\psi D\tilde{a}\ \exp\bigg(iS_{fermion}[\psi ; \tilde{a}]
+\frac{i}{2}S_{CS}[\tilde{a}]-iS_{BF}[\tilde{a} ;A]+iS_{CS}[A]\bigg)
\nn\\&=&\int Da\ \exp\bigg(iS_{BF}[a ; A]\bigg)\int Df\ Z_{scalar}[a+df]
\eea
by computing:
\bea
-iS_{BF}[\tilde{a}; a]&=&-iS_{BF}[\tilde{a}; A-\tilde{a}+dg]=-iS_{BF}[\tilde{a}; A]+2iS_{CS}[\tilde{a}],
\nn\\
iS_{BF}[a; A]&=&iS_{BF}[A-\tilde{a}+dg; A]=2iS_{CS}[A]-iS_{BF}[\tilde{a}; A],
\nn\\
-iS_{CS}[a]&=&-iS_{CS}[A-\tilde{a}+dg]=-iS_{CS}[A]-iS_{CS}[\tilde{a}]+iS_{BF}[A; \tilde{a}-dg]
\nn\\
&=&-iS_{CS}[A]-iS_{CS}[\tilde{a}]+iS_{BF}[A; \tilde{a}],
\eea
\bea
-\frac{i}{2}S_{CS}[\tilde{a}]-iS_{BF}[\tilde{a}, a]-iS_{CS}[a]+iS_{BF}[a; A]&=&\frac{i}{2}S_{CS}[\tilde{a}]-iS_{BF}[\tilde{a}; A]+iS_{CS}[A]
\nn\\
\eea
through:
\bea
S_{BF}[\tilde{a}, \tilde{a}]&=&-S_{BF}[-\tilde{a}, \tilde{a}]=-S_{BF}[\tilde{a}, -\tilde{a}]=2S_{CS}[\tilde{a}],
\nn\\
S_{BF}[\tilde{a}, A+B]&=&S_{BF}[\tilde{a}, A]+S_{BF}[\tilde{a}, B],
\nn\\
S_{CS}[A]&=&S_{CS}[-A],
\nn\\
S_{CS}[A+B]&=&S_{CS}[A]+S_{CS}[B]-S_{BF}[A ;B].
\eea
Then we could use the time reversed partition function
\bea
&&\int D\psi DA\ \exp\bigg(iS_{fermion}[\psi ; A]+\frac{i}{2}S_{CS}[A]-iS_{BF}[A ;C]\bigg)
\nn\\
&=&e^{-iS_{CS}[C]}\int Df\ Z_{scalar}[-C+df].
\nn\\
\eea
Therefore, we obtain
\bea
\int Df\ Z_{scalar}[-A+df]=\int Da\ \exp\bigg(iS_{BF}[a ; A]\bigg)\int Df\ Z_{scalar}[a+df].
\eea
The duality is a little complicated because we include the path integer of $f$. Since the gauge transformation of $a$ is $d\lambda$ and the shifting of $a$, $a\rightarrow a-df$, does not modify the $BF$ term, we could rewrite the duality
\bea
\int Df\ Z_{scalar}[-A+df]\sim\int Da\ \exp\bigg(iS_{BF}[a ; A]\bigg)Z_{scalar}[a].
\eea
Now we could clearly find the effects of the path integral of $f$. Fortunately, the path integral of $f$ could be absorbed by $\phi\rightarrow\phi\cdot\exp(-if)$. Thus, we could show that the boson-boson duality is not affected by the path integral of $f$:
\bea
Z_{scalar}[-A]\sim\int Da\ \exp\bigg(iS_{BF}[a ; A]\bigg)Z_{scalar}[a].
\eea
We could find that the interacting scalar field theory with the background gauge field could dual to the interacting scalar field theory with the dynamical gauge field couple to the $BF$ term. One interesting application of the duality is that the entanglement entropy is hard to define when a gauge field appears, but the entanglement entropy in the interacting scalar field theory with the background gauge field could be defined easily. Although the duality is only guaranteed at an IR limit, we could use a renormalization group flow to obtain the entanglement entropy at all energy scales. Thus, we think that the entanglement entropy could be defined at all energy scales when dynamical gauge fields appear. Even if Hilbert spaces of two theories are the same, we possibly have a non-local transformation between two Hilbert spaces. Thus, they have complicated equivalence of the universal term of the entanglement entropy by choosing some basis through a non-local transformation between two Hilbert spaces.

\subsection{Fermion-Fermion Duality}
We derive a fermion-fermion duality from the conjecture
\bea
&&Z_{fermion}[C]
\nn\\
&=&\int D\phi D\tilde{a}\ \exp\big(i S_{scalar}[\phi; \tilde{a}]+iS_{CS}[\tilde{a}]+iS_{BF}[\tilde{a}; C]\big)\exp\bigg(\frac{i}{2}S_{CS}[C]\bigg),
\nn\\
\eea
where $C\equiv 2a+A$, because we want that our theory is gauge invariant with a spin structure and results of integration of the dynamical gauge fields should obey the Dirac quantization condition, which is $\int dC=2\pi$. Then we could add the $BF$ coupling to obtain
\bea
&&\int Da\ Z_{fermion}[C]\cdot\exp\bigg(\frac{i}{2}S_{BF}[C; A]\bigg)
\nn\\
&=&\int D\phi D\tilde{a}Da\ \exp\big(i S_{scalar}[\phi; \tilde{a}]+iS_{CS}[\tilde{a}]+iS_{BF}[\tilde{a}; C]\big)\exp\bigg(\frac{i}{2}S_{CS}[C]\bigg)
\nn\\
&&\times\exp\bigg(\frac{i}{2}S_{BF}[C; A]\bigg).
\nn\\
\eea
We could integrate $a$ out in the boson theory to obtain $dC=-(dA+2d\tilde{a})$, which is equivalent to $C=-A-2\tilde{a}+dg$. Thus, we get
\bea
&&\int D\phi D\tilde{a}Da\ \exp\big(i S_{scalar}[\phi; \tilde{a}]+iS_{CS}[\tilde{a}]+iS_{BF}[\tilde{a}; C]\big)\exp\bigg(\frac{i}{2}S_{CS}[C]\bigg)
\nn\\
&&\times\exp\bigg(\frac{i}{2}S_{BF}[C; A]\bigg)
\nn\\
&=&\int D\phi D\tilde{a}\ \exp\big(i S_{scalar}[\phi; \tilde{a}]-iS_{CS}[\tilde{a}]-iS_{BF}[\tilde{a}; A]-\frac{i}{2}S_{CS}[A]\big)
\nn\\
\eea
by computing:
\bea
iS_{BF}[\tilde{a}; C]&=&iS_{BF}[\tilde{a}, -A-2\tilde{a}+dg]=-iS_{BF}[\tilde{a}, A]-4S_{CS}[\tilde{a}],
\nn\\
\frac{i}{2}S_{CS}[C]&=&\frac{i}{2}S_{CS}[-A-2\tilde{a}+dg]=\frac{i}{2}S_{CS}[A]+2iS_{CS}[\tilde{a}]+iS_{BF}[\tilde{a}; A],
\nn\\
\frac{i}{2}S_{BF}[C; A]&=&\frac{i}{2}S_{BF}[-A-2\tilde{a}+dg, A]=-iS_{CS}[A]-iS_{BF}[\tilde{a}; A],
\eea
\bea
iS_{CS}[\tilde{a}]+iS_{BF}[\tilde{a}, C]+\frac{i}{2}S_{CS}[C]+\frac{i}{2}S_{BF}[C; A]=-iS_{CS}[\tilde{a}]-iS_{BF}[\tilde{a}; A]-\frac{i}{2}S_{CS}[A].
\nn\\
\eea
Hence, we could use
\bea
Z_{fermion}[A]\cdot\exp\bigg(\frac{i}{2}S_{CS}[A]\bigg)=\int D\phi Da\ \exp\bigg(iS_{scalar}[\phi; a]-iS_{CS}[a]-iS_{BF}[a;A]\bigg)
\nn\\
\eea
to obtain:
\bea
Z_{fermion}[A]&=&\int D\psi Da\ \exp\bigg(i S_{fermion}[\psi; 2a+A]+\frac{i}{2}S_{BF}[2a+A; A]\bigg)
\nn\\
&=&
\int D\psi Da\ \exp\bigg(i S_{fermion}[\psi; 2a+A]+iS_{BF}[a; A]+iS_{CS}[A]\bigg).
\nn\\
\eea
We could find that a non-interacting fermion theory could dual to an interacting fermion theory with the dynamical gauge fields without affecting by the non-trivial holonomy. 

\subsection{Self-Dual Fermions}
The partition function of the self-dual fermions is 
\bea
Z_{sdf}=\int Da\ Z_{fermion}[a+C]Z_{fermion}[a-C]\cdot\exp\bigg(iS_{BF}[a; A]\bigg).
\eea
The meaning of the self-dual is that physics is not changed by exchanging $A$ and $C$. We derive the self-dual from the conjecture of the duality between the boson system and the fermion system in three dimensions. The conjecture of the duality between a boson system and a fermion system in three dimensions at an IR limit could be written as
\bea
&&Z_{fermion}[A_1]Z_{fermion}[A_2]
\nn\\
&=&\int D\phi Da\ \exp\bigg(iS_{scalar}[\phi; a]+iS_{CS}[a]+iS_{BF}[a; A_1]\bigg)
\nn\\
&&\times \int D\tilde{\phi}D\tilde{a}\ \exp\bigg(+iS_{scalar}[\tilde{\phi}; \tilde{a}]-iS_{CS}[\tilde{a}]-iS_{BF}[\tilde{a}; A_2]\bigg)
\nn\\
&&\times\exp\bigg(\frac{i}{2}S_{CS}[A_1]-\frac{i}{2}S_{CS}[A_2]\bigg).
\nn\\
\eea
The conjecture combines the partition function of the non-interacting fermion theory and the time reversed partition function of the non-interacting fermion theory. Thus, the partition function of the self-dual fermions could be rewritten as
\bea
Z_{sdf}&=&\int DaD\phi_1D\tilde{a}_1D\phi_2D\tilde{a}_2
\nn\\
&&\times\exp\bigg(iS_{scalar}[\phi_1;\tilde{a}_1]+iS_{scalar}[\phi_2; \tilde{a}_2]+iS_{CS}[\tilde{a_1}]-iS_{CS}[\tilde{a_2}]
\nn\\
&&+iS_{BF}[\tilde{a_1}-\tilde{a}_2; a]+iS_{BF}[\tilde{a_1}+\tilde{a}_2; C]+iS_{BF}[a; A+C]\bigg)
\nn\\
\eea
by computing:
\bea
&&S_{BF}[\tilde{a_1}; a+C]-S_{BF}[\tilde{a}_2; a-C]=S_{BF}[\tilde{a}_1-\tilde{a}_2; a]+S_{BF}[\tilde{a}_1+\tilde{a}_2;C],
\nn\\
&&\frac{1}{2}S_{CS}[a+C]-\frac{1}{2}S_{CS}[a-C]+S_{BF}[a; A]=S_{BF}[a;C]+S_{BF}[a; A]=S_{BF}[a; A+C].
\nn\\
\eea
Now we integrate $a$ out, then it is equivalently using 
\bea
d\tilde{a}_1-d\tilde{a}_2+dA+dC=0
\eea
or
\bea
\tilde{a}_1-\tilde{a}_2=-A-C+dg.
\eea
Hence, the partition function of the self-dual fermions becomes:
\bea
Z_{sdf}&\sim&\int D\phi_1D\phi_2D c_+Dg\ \exp\bigg(iS[\phi_1, \phi_2, c_+; A+C-dg]-\frac{i}{2}S_{BF}[c_+; A-C]\bigg)
\nn\\
&\sim&
\int D\phi_1D\phi_2D c_+\ \exp\bigg(iS[\phi_1, \phi_2, c_+; A+C]-\frac{i}{2}S_{BF}[c_+; A-C]\bigg)
\eea
by computing:
\bea
&&
S_{BF}[\tilde{a}_1-\tilde{a_2}; a]+S_{BF}[\tilde{a_1}+\tilde{a}_2; C]+S_{BF}[a; A+C]
\nn\\
&=& S_{BF}[-A-C+dg; a]+S_{BF}[c_+; C]+S_{BF}[a; A+C]=S_{BF}[c_+; C],
\nn\\
&&
S_{CS}[\tilde{a}_1]-S_{CS}[\tilde{a}_2]=S_{CS}[\frac{1}{2}c_++\frac{1}{2}c_-]-S_{CS}[\frac{1}{2}c_+ -\frac{1}{2}c_-]
=\frac{1}{2}S_{BF}[c_+; c_-]
\nn\\
&=&\frac{1}{2}S_{BF}[c_+; -A-C+dg]=\frac{1}{2}S_{BF}[c_+; -A-C],
\nn\\
&&S_{CS}[\tilde{a}_1]-S_{CS}[\tilde{a}_2]+S_{BF}[\tilde{a}_1-\tilde{a_2}; a]+S_{BF}[\tilde{a_1}+\tilde{a}_2; C]+S_{BF}[a; A+C]
\nn\\
&=&\frac{1}{2}S_{BF}[c_+; -A-C]+S_{BF}[c_+; C]=S_{BF}[c_+;-\frac{A}{2}+\frac{C}{2}]=-\frac{1}{2}S_{BF}[c_+; A-C],
\nn\\
\eea
where
\bea
c_+\equiv \tilde{a}_1+\tilde{a}_2, \qquad c_-\equiv\tilde{a}_1-\tilde{a}_2.
\eea
We could remove $dg$ by using the a field redefinition of the scalar fields.
Therefore, if we exchange $A$ and $C$ in the partition function of the self-dual fermions, we could obtain the time reversed partition function of the self-dual fermions.

\subsection{Self-Dual Bosons}
We start from the partition function of the self-dual bosons and the conjecture:
\bea
Z_{sdb}&=&\int Da\ Z_{scalar}[a+C]Z_{scalar}[a-C]\cdot\exp\bigg(iS_{BF}[a; A]\bigg)
\nn\\
&=&\int Da D\psi_1DA_1\ \exp\bigg(iS_{fermion}[\psi_1; A_1]-\frac{i}{2}S_{CS}[A_1]-iS_{BF}[A_1; a+C]\bigg)
\nn\\
&&\times\int D\psi_2DA_2\ \exp\bigg(iS_{fermion}[\psi_2; A_2]+\frac{i}{2}S_{CS}[A_2]+iS_{BF}[A_2; a-C]\bigg)
\nn\\
&&\times\exp\bigg(-iS_{CS}[a+C]+iS_{CS}[a-C]+iS_{BF}[a; A]\bigg)
\nn\\
&=&\int Da D\psi_1DA_1\ \exp\bigg(iS_{fermion}[\psi_1; A_1]-\frac{i}{2}S_{CS}[A_1]-iS_{BF}[A_1; a+C]\bigg)
\nn\\
&&\times\int D\psi_2DA_2\ \exp\bigg(iS_{fermion}[\psi_2; A_2]+\frac{i}{2}S_{CS}[A_2]+iS_{BF}[A_2; a-C]\bigg)
\nn\\
&&\times\exp\bigg(iS_{BF}[a; A-2C]\bigg).
\eea
The last equality is obtained by computing:
\bea
-iS_{CS}[a+C]&=&-iS_{CS}[a]-iS_{CS}[C]-iS_{BF}[a; C],
\nn\\
iS_{CS}[a-C]&=&iS_{CS}[a]+iS_{CS}[C]-iS_{BF}[a; C],
\eea
\bea
-iS_{CS}[a+C]+iS_{CS}[a-C]+iS_{BF}[a; A]=iS_{BF}[a; A-2C].
\eea
Now we integrate $a$ out in the fermion theory, then it is equivalent to using
\bea
dA_1-dA_2=dA-2dC
\eea
or
\bea
A_1-A_2=A-2C+dg.
\eea
Then we could obtain:
\bea
Z_{sdb}&\sim&\int D\psi_1D\psi_2Dc_+Dg\ \exp\bigg(i\tilde{S}[\psi_1, \psi_2, c_+; A-2C+dg]-\frac{i}{4}S_{BF}[c_+; A+2C]\bigg)
\nn\\
&\sim&\int D\psi_1D\psi_2Dc_+\ \exp\bigg(i\tilde{S}[\psi_1, \psi_2, c_+; A-2C]-\frac{i}{4}S_{BF}[c_+; A+2C]\bigg)
\eea
by computing:
\bea
-\frac{i}{2}S_{CS}[A_1]&=&-\frac{i}{2}S_{CS}[\frac{1}{2}c_++\frac{1}{2}c_-]=-\frac{i}{8}S_{CS}[c_+]-\frac{i}{8}S_{CS}[c_-]-\frac{i}{8}S_{BF}[c_+; c_-],
\nn\\
\frac{i}{2}S_{CS}[A_2]&=&\frac{i}{2}S_{CS}[\frac{1}{2}c_+-\frac{1}{2}c_-]=\frac{i}{8}S_{CS}[c_+]+\frac{i}{8}S_{CS}[c_-]-\frac{i}{8}S_{BF}[c_+; c_-],
\nn\\
-iS_{BF}[A_1; a+C]&=&-iS_{BF}[\frac{1}{2}c_++\frac{1}{2}c_-;a+C]
\nn\\
&=&-\frac{i}{2}S_{BF}[c_+;a]-\frac{i}{2}S_{BF}[c_+;C]-\frac{i}{2}S_{BF}[c_-; a]-\frac{i}{2}S_{BF}[c_-; C],
\nn\\
iS_{BF}[A_2; a-C]&=&iS_{BF}[\frac{1}{2}c_+-\frac{1}{2}c_-;a-C]
\nn\\
&=&\frac{i}{2}S_{BF}[c_+;a]-\frac{i}{2}S_{BF}[c_+;C]-\frac{i}{2}S_{BF}[c_-; a]+\frac{i}{2}S_{BF}[c_-; C],
\nn\\
iS_{BF}[a; A-2C]&=&iS_{BF}[a; A]-2iS_{BF}[a; C],
\eea
\bea
&&-\frac{i}{2}S_{CS}[A_1]+\frac{i}{2}S_{CS}[A_2]-iS_{BF}[A_1; a+C]+iS_{BF}[A_2; a-C]+iS_{BF}[a; A-2C]
\nn\\
&=&-\frac{i}{4}S_{BF}[c_+; A-2C+dg]-iS_{BF}[c_+; C]-iS_{BF}[A-2C+dg; a]
\nn\\
&&+iS_{BF}[a; A]+iS_{BF}[a;-2C]
\nn\\
&=&-\frac{i}{4}S_{BF}[c_+; A-2C+dg]-iS_{BF}[c_+; C]=-\frac{i}{4}S_{BF}[c_+; A+2C],
\eea
where $c_+\equiv A_1+A_2$ and $c_-\equiv A_1-A_2$.
We could remove $dg$ by doing a field redefinition of the fermion fields. Now we obtain that the partition function of the self-dual bosons is equivalent to the time reversed partition function of the self-dual bosons by exchanging $A$ and $-2C$.
\subsection{A Vortex-Vortex Duality in Three Dimensions}
A vortex-vortex duality is a mapping between  monopole operators. We start from the conjecture
\bea
&&Z_{scalar}[C]e^{imS_{CS}[C]}
\nn\\
&=&e^{i(m-1)S_{CS}[C]}\int D\psi DA\ \exp\bigg(i S_{fermion}[\psi; A]-\frac{i}{2}S_{CS}[A]-iS_{BF}[A; C]\bigg),
\nn\\
\eea
where $m$ is a constant. Thus, we could rewrite the conjecture:
\bea
&&\int Da\ Z_{scalar}[a]\exp\bigg(imS_{CS}[a]+iS_{BF}[a; A]\bigg)
\nn\\
&=&
\int Da DC\ Z_{fermion}[a]
\nn\\
&&\times\exp\bigg(-\frac{i}{2}S_{CS}[a]-iS_{BF}[a; C]+i(m-1)S_{CS}[C]+iS_{BF}[C; A]\bigg)
\nn\\
&=&
\int Da\ Z_{fermion}[a]
\nn\\
&&\times\exp\bigg(-\frac{i}{2}S_{CS}[a]-\frac{i}{m-1}S_{BF}[a; a-A+dg]+\frac{i}{m-1}S_{CS}[a-A+dg]
\nn\\
&&+\frac{i}{m-1}S_{BF}[a-A+dg; A]\bigg).
\nn\\
\eea
The last identity is obtained because the integration of $a$ in the fermion theory is equivalent to using 
\bea
(m-1)dC=da-dA
\eea
or
\bea
(m-1)C=a-A+dg.
\eea
If $m\neq 2$, the integration of $a$ could not be performed because the result of the integration does not satisfy the Dirac quantization condition. In other words, the function $g$ should not be globally defined or $ddg\neq 0$. Now we set $m=2$ because we want to focus on a globally defined function for $g$. Finally, we could obtain:
\bea
&&\int Da\ Z_{scalar}[a]\exp\bigg(2iS_{CS}[a]+iS_{BF}[a; A]\bigg)
\nn\\
&=&\int Da\ Z_{fermion}[a]
\nn\\
&&\times\exp\bigg(-\frac{i}{2}S_{CS}[a]-iS_{BF}[a; a-A]+iS_{CS}[a-A]+iS_{BF}[a-A; A]\bigg).
\nn\\
&=&\int Da\ Z_{fermion}[a]\exp\bigg( -\frac{3i}{2}S_{CS}[a]+iS_{BF}[a; A]-iS_{CS}[A]\bigg)
\eea
by computing:
\bea
-iS_{BF}[a; a-A]&=&-2iS_{CS}[a]+iS_{BF}[a; A],
\nn\\
iS_{CS}[a-A]&=&iS_{CS}[a]+iS_{CS}[A]-iS_{BF}[a; A],
\nn\\
iS_{BF}[a-A; A]&=&iS_{BF}[a; A]-2iS_{CS}[A],
\eea
\bea
&&-\frac{i}{2}S_{CS}[a]-iS_{BF}[a; a-A]+iS_{CS}[a-A]+iS_{BF}[a-A; A]
\nn\\
&=&-\frac{3i}{2}S_{CS}[a]+iS_{BF}[a; A]-iS_{CS}[A].
\nn\\
\eea
The result is that the partition function of the scalar field theory coupled to the Chern-Simons gauge theory with the level two and the BF term is equivalent to the partition function of a fermion theory coupled to the Chern-Simons gauge theory with the level -3/2, BF term and the Chern-Simons gauge theory with the level -1, in which the gauge field is a background gauge field.

If we turn off the background gauge field $A$, we could compute the equation of the motion for $a^0$ from the boson theory to get
\bea
\rho_{scalar}=-\frac{f}{\pi},\qquad f\equiv \partial_1a_2-\partial_2a_1.
\eea
If a monopole is generated, we need to have two scalar modes. The lowest excited state of the scalar field has spin $1/2$ so the two scalar modes could give a monopole with spin one. If we turn off the background gauge field $A$, we compute the equation of motion of $a^0$ from the fermion theory to obtain
\bea
\rho_{fermion}=\frac{3f}{4\pi}.
\eea
If a monopole is generated, single monopole has charge $3/2$. The zero mode could have charge $1/2$ and gives spin zero monopole. The charge $3/2$ monopole or the first excited state also has spin one.

\section{The Mean-Field Study of the Duality in Three Dimensions}
\label{3}
We use a mean-field method to study the duality \cite{Ma:2016yas, Karch:2016aux} between the boson systems and the fermion systems at an IR limit. The massless Dirac fermion could be obtained from
\bea
S_{gn}=\int d^3x\ \bigg(\bar{\psi}\gamma^{\mu}\big(i\partial_{\mu}-A_{\mu}\big)\psi+\frac{g^2}{2}\big(\bar{\psi}\psi\big)^2\bigg)
\eea
at an IR limit or $g^2\rightarrow 0$. We rewrite the action 
\bea
\int d^3x\ \bigg(\bar{\psi}\gamma^{\mu}\big(i\partial_{\mu}-A_{\mu}\big)\psi-\frac{1}{2g^2}m^2-m\bar{\psi}\psi\bigg)
\eea
by introducing the auxiliary field $m$. When we take the IR limit, the auxiliary field $m$ approaches to a constant or a fixed value from the mean-field analysis. Thus, we could get a new fermionic mass term
\bea
\int d^3x\ \bigg(\bar{\psi}\gamma^{\mu}\big(i\partial_{\mu}-A_{\mu}\big)\psi-m\bar{\psi}\psi\bigg).
\eea
Indeed, this mass term should not affect the dynamics because $m\sim\bar{\psi}\psi$ is a constant. Hence, the fermion theory is
\bea
Z_{fermion}[A]\cdot e^{-\frac{i}{2}S_{CS}[A]}\delta\big(\bar{\psi}\psi-\hat{m}\big),
\eea
where $\hat{m}$ is a constant, in the conjecture of the duality between the boson system and fermion system.

Now we consider the scalar field theory
\bea
\int d^3x\ \bigg(|(\partial_{\mu}+ia_{\mu})\phi|^2-\alpha|\phi|^4\bigg)\rightarrow \int d^3x\ \bigg(|(\partial_{\mu}+ia_{\mu})\phi|^2-\sigma|\phi|^2+\frac{1}{4\alpha}\sigma^2\bigg)
\eea
by introducing the auxiliary field $\sigma$. To understand the mean-field analysis of the scalar field theory, we use the conjecture
\bea
&&\lim_{g\rightarrow 0,\alpha\rightarrow\infty}\int D\mu D\sigma\ \Bigg\{\Bigg\lbrack\ \int D\psi\ \exp\bigg\lbrack i\int d^3x\ \bigg(\bar{\psi}\gamma^{\mu}\big(i\partial_{\mu}-A_{\mu}\big)\psi-\frac{1}{2g^2}\sigma^2-\sigma\bar{\psi}\psi\bigg)\bigg\rbrack
\nn\\
&&\times \exp\bigg(-\frac{i}{2}S_{CS}[A]+i\mu( \sigma-\tilde{m})\bigg)\Bigg\rbrack
\nn\\
&-&\Bigg\lbrack\int D\phi Da\ \exp\bigg\lbrack i\int d^3x\ \bigg(|(\partial_{\mu}+ia_{\mu})\phi|^2-\sigma|\phi|^2+\frac{1}{4\alpha}\sigma^2\bigg)\bigg\rbrack
\nn\\
&&\times\exp\bigg(iS_{CS}[a]+iS_{BF}[a; A]+i\mu(\sigma-\tilde{m})\bigg)\Bigg\rbrack\Bigg\}\approx 0,
\nn\\
\eea
where $\tilde{m}$ is determined by the mean-field analysis of the fermion field.

We first integrate $\mu$ out, then we could obtain
\bea
&&\lim_{g\rightarrow 0,\alpha\rightarrow\infty}\ \Bigg\lbrack\ \int D\psi\ \exp\bigg\lbrack i\int d^3x\ \bigg(\bar{\psi}\gamma^{\mu}\big(i\partial_{\mu}-A_{\mu}\big)\psi-\frac{1}{2g^2}\tilde{m}^2-\tilde{m}\bar{\psi}\psi\bigg)\bigg\rbrack
\nn\\
&&\times \exp\bigg(-\frac{i}{2}S_{CS}[A]\bigg)\Bigg\rbrack
\nn\\
&\approx&\Bigg\lbrack\int D\phi Da\ \exp\bigg\lbrack i\int d^3x\ \bigg(|(\partial_{\mu}+ia_{\mu})\phi|^2-\tilde{m}|\phi|^2+\frac{1}{4\alpha}\tilde{m}^2\bigg)\bigg\rbrack
\nn\\
&&\times
\exp\bigg(iS_{CS}[a]+iS_{BF}[a; A]\bigg)\Bigg\rbrack.
\nn\\
\eea
After we integrate the fermion field out, we could use the mean-field analysis to determine the constant $\tilde{m}$. Here, we expect that if two partition functions are the same after we sum over all configurations of $\tilde{m}$. Thus, the conjecture is equivalent to
\bea
&&\lim_{\alpha\rightarrow\infty}\ \Bigg\lbrack\ \int D\psi\ \exp\bigg\lbrack i\int d^3x\ \bigg(\bar{\psi}\gamma^{\mu}\big(i\partial_{\mu}-A_{\mu}\big)\psi\bigg)\bigg\rbrack
\nn\\
&&\times \exp\bigg(-\frac{i}{2}S_{CS}[A]\bigg)\Bigg\rbrack\delta\big(\bar{\psi}\psi-\hat{m}\big)
\nn\\
&\approx&\Bigg\lbrack\int D\phi Da\ \exp\bigg\lbrack i\int d^3x\ \bigg(|(\partial_{\mu}+ia_{\mu})\phi|^2-\tilde{m}|\phi|^2+\frac{1}{4\alpha}\tilde{m}^2\bigg)\bigg\rbrack
\nn\\
&&\times
\exp\bigg(iS_{CS}[a]+iS_{BF}[a; A]\bigg)\Bigg\rbrack,
\nn\\
\eea
where $\hat{m}=-\tilde{m}/g^2$ is also determined form the mean-field analysis at the limit $g\rightarrow 0$.
Now we take $\alpha\rightarrow\infty$, the conjecture becomes
\bea
&&\int D\psi\ \exp\bigg\lbrack i\int d^3x\ \bigg(\bar{\psi}\gamma^{\mu}\big(i\partial_{\mu}-A_{\mu}\big)\psi\bigg)\bigg\rbrack
\cdot \exp\bigg(-\frac{i}{2}S_{CS}[A]\bigg)\delta\big(\bar{\psi}\psi-\hat{m}\big)
\nn\\
&\approx&\int D\phi Da\ \exp\bigg\lbrack i\int d^3x\ \bigg(|(\partial_{\mu}+ia_{\mu})\phi|^2-\tilde{m}|\phi|^2\bigg)\bigg\rbrack\cdot
\exp\bigg(iS_{CS}[a]+iS_{BF}[a; A]\bigg).
\nn\\
\eea

Our result is that the fermion theory with the constant condensation could dual to the massive scalar field theory at the IR limit. Because we consider the IR limit, the kinetic term of the dynamical gauge field does not appear to affect the validity of the conjecture.

\section{Duality at a Finite Temperature in Three Dimensions}
\label{4}
To build the duality at a finite temperature \cite{Ma:2016yas}, we consider a one-loop effective potential of the Dirac fermion at a finite temperature to obtain
\bea
\exp\bigg(-\frac{1}{2}S_{CST}[A; T]\bigg)\equiv\exp\bigg\lbrack i\frac{1}{2}\frac{1}{4\pi}\frac{m}{|m|}\tanh\bigg(\frac{|m|}{2T}\bigg)\int_0^{\frac{1}{T}}dt\int d^2x\ \epsilon_{\mu\nu\rho}A_{\mu}\partial_{\nu}A_{\rho}\Bigg\rbrack
\eea
in an Euclidean space.
The action of the scalar field theory at a finite temperature is
\bea
S_{scalarT}[\phi; A; T]=\int_0^{\frac{1}{T}}\int d^2x\ \bigg(|\partial_{\mu}+iA_{\mu}|^2+m_b^2|\phi|^2+\lambda|\phi|^4\bigg)
\eea
in the Euclidean space. The partition function of the Dirac fermion theory at a finite temperature is
\bea
Z_{fermionT}[A; T]\equiv\int D\psi\exp\bigg\lbrack\int_0^{\frac{1}{T}}\int d^2x\ \bigg(\bar{\psi}\gamma_{\mu}\big(i\partial_{\mu}-A_{\mu}\big)-m_f\bar{\psi}\psi\bigg)\bigg\rbrack
\eea
in the Euclidean space. Thus, we propose that the suitable conjecture should be written as
\bea
&&Z_{fermionT}[A; T]\cdot\exp\bigg(-\frac{1}{2}S_{CST}[A; T]\bigg)
\nn\\
&=&\int D\phi Da\exp\bigg(-S_{scalarT}[\phi; a; T]-S_{CST}[a+A; T]+S_{CST}[A; T]\bigg).
\eea
When the fermion mass is positive, the theory is trivial and this case also corresponds to the positive boson mass. Alternatively, we turn on the negative fermion mass and obtain the Chern-Simons term with level one. This case could correspond to the negative boson mass. The correspondence is not only valid at a low temperature, but also at a high temperature. When a temperature is high enough, the effective theory of the fermion theory and the boson theory are also trivial. Thus, we want to argue that the duality is valid for each temperature. Indeed, we do not only use a Wick rotation to define the conjecture. The coefficient of the induced Chern-Simons term at a finite temperature is also related to a temperature. This means that the equivalence of the bosonic system and the fermionic system at a finite temperature is not trivial.

When the Chern-Simons theory has a temperature dependent coefficient, the gauge symmetry is broken. Thus, our proposal of the duality loses the gauge symmetry. However, the gauge symmetry could be restored by resumming to all orders. We need to remind that the proposal is not guaranteed when we consider higher orders of the resummation. However, we expect that the duality could be found when we use an effective theory (by resumming to all orders) to do perturbation at the leading order of the effective theory.

\section{Discussion of a Duality Web in Odd Dimensions}
\label{5}
We discuss a duality web in $2p+1$ dimensions by using an $SL(2)$ transformation of the non-interacting $p$-form theory with a theta term in $2p+2$ dimensions and a boundary theory in $2p+1$ dimensions. When $p=1$, the $SL(2)$ transformations \cite{Witten:2003ya} in four dimensions could be related to a three dimensional duality web \cite{Seiberg:2016gmd}. From the four dimensional perspective or the bulk perspective, the result does not totally depend on the conjecture. Thus, the combination of a bulk theory and a boundary theory is a consistent study in the three dimensional duality web. We want to generalize the study to $2p+2$ dimensions to understand constraints and difficulties of a duality web in other odd dimensions.

\subsection{Electric-Magnetic Duality of the $p$-form Theory in $2p+2$ Dimensions}
The action of the non-interacting $p$-form theory with a theta term in $2p+2$ dimensions is 
\bea
S_{p}&=&\int d^{2p+2}x\ \bigg(\frac{1}{2e^2(p+1)!}F_{\mu_1\mu_2\cdots\mu_{p+1}}F^{\mu_1\mu_2\cdots\mu_{p+1}}
\nn\\
&&+\frac{\theta}{8\big((p+1)!\big)^2\pi^2}\epsilon^{\mu_1\mu_2\cdots\mu_{p+1}\nu_1\nu_2\cdots\nu_{p+1}}F_{\mu_1\mu_2\cdots\mu_{p+1}}F_{\nu_1\nu_2\cdots\nu_{p+1}}
\bigg),
\nn\\
\eea
where $F\equiv dA$, $A$ is the $p$-form potential, and $F$ is the field strength associated to the $p$-form potential. The action could also be written as
\bea
S_p=\frac{i}{8\pi}\int d^{2p+2}x\ \bigg(\tau^*F^+_{\mu_1\mu_2\cdots\mu_{p+1}}F^{+,\ \mu_1\mu_2\cdots\mu_{p+1}}
-\tau F^-_{\mu_1\mu_2\cdots\mu_{p+1}}F^{-,\ \mu_1\mu_2\cdots\mu_{p+1}}\bigg)
\eea
where 
\bea
F^{\pm}_{\mu_1\mu_2\cdots\mu_{p+1}}&\equiv&\sqrt{\frac{1}{2(p+1)!}}\bigg(F_{\mu_1\mu_2\cdots\mu_{p+1}}\pm\frac{i}{(p+1)!}\epsilon_{\mu_1\mu_2\cdots\mu_{p+1}\nu_1\nu_2\cdots\nu_{p+1}}F^{\nu_1\nu_2\cdots\nu_{p+1}}\bigg),
\nn\\
\tau&\equiv&-\frac{\theta}{2\pi}+\frac{2\pi i}{e^2}.
\eea
The first term of $S_p$ could be written as:
\bea
&&\tau^*F^+_{\mu_1\mu_2\cdots\mu_{p+1}}F^{+,\ \mu_1\mu_2\cdots\mu_{p+1}}
\nn\\
&=&\bigg(-\frac{\theta}{2\pi}-\frac{2\pi i}{e^2}\bigg)\frac{1}{2(p+1)!}\bigg(F_{\mu_1\mu_2\cdots\mu_{p+1}}+\frac{i}{(p+1)!}\epsilon_{\mu_1\mu_2\cdots\mu_{p+1}\nu_1\nu_2\cdots\nu_{p+1}}F^{\nu_1\nu_2\cdots\nu_{p+1}}\bigg)
\nn\\
&&\times \bigg(F^{\mu_1\mu_2\cdots\mu_{p+1}}+\frac{i}{(p+1)!}\epsilon^{\mu_1\mu_2\cdots\mu_{p+1}\nu_1\nu_2\cdots\nu_{p+1}}F_{\nu_1\nu_2\cdots\nu_{p+1}}\bigg)
\nn\\
&=&\frac{1}{2(p+1)!}\bigg(-\frac{\theta}{2\pi}-\frac{2\pi i}{e^2}\bigg)\bigg(F_{\mu_1\mu_2\cdots\mu_{p+1}}F^{\mu_1\mu_2\cdots\mu_{p+1}}
+F_{\mu_1\mu_2\cdots\mu_{p+1}}F^{\mu_1\mu_2\cdots\mu_{p+1}}
\nn\\
&&+\frac{2i}{(p+1)!}\epsilon_{\mu_1\mu_2\cdots\mu_{p+1}\nu_1\nu_2\cdots\nu_{p+1}}
F^{\mu_1\mu_2\cdots\mu_{p+1}}F^{\nu_1\nu_2\cdots\nu_{p+1}}\bigg)
\nn\\
&=&\frac{1}{(p+1)!}\bigg(-\frac{\theta}{2\pi}-\frac{2\pi i}{e^2}\bigg)F_{\mu_1\mu_2\cdots\mu_{p+1}}F^{\mu_1\mu_2\cdots\mu_{p+1}}
\nn\\
&&+\frac{1}{\big((p+1)!\big)^2}\bigg(\frac{2\pi}{e^2}-\frac{i\theta}{2\pi}\bigg)\epsilon_{\mu_1\mu_2\cdots\mu_{p+1}\nu_1\nu_2\cdots\nu_{p+1}}
F^{\mu_1\mu_2\cdots\mu_{p+1}}F^{\nu_1\nu_2\cdots\nu_{p+1}}.
\eea
Thus, we could obtain:
\bea
&&\tau^*F^+_{\mu_1\mu_2\cdots\mu_{p+1}}F^{+,\ \mu_1\mu_2\cdots\mu_{p+1}}-\tau F^-_{\mu_1\mu_2\cdots\mu_{p+1}}F^{-,\ \mu_1\mu_2\cdots\mu_{p+1}}
\nn\\
&=&-\frac{4\pi i}{e^2}\frac{1}{(p+1)!}F_{\mu_1\mu_2\cdots\mu_{p+1}}F^{\mu_1\mu_2\cdots\mu_{p+1}}
\nn\\
&&-\frac{i\theta}{\pi}\frac{1}{\big((p+1)!\big)^2}\epsilon_{\mu_1\mu_2\cdots\mu_{p+1}\nu_1\nu_2\cdots\nu_{p+1}}
F^{\mu_1\mu_2\cdots\mu_{p+1}}F^{\nu_1\nu_2\cdots\nu_{p+1}},
\nn\\
S_p&=&\frac{i}{8\pi}\int d^{2p+2}x\ \bigg(\tau^*F^+_{\mu_1\mu_2\cdots\mu_{p+1}}F^{+,\ \mu_1\mu_2\cdots\mu_{p+1}}
-\tau F^-_{\mu_1\mu_2\cdots\mu_{p+1}}F^{-,\ \mu_1\mu_2\cdots\mu_{p+1}}\bigg)
\nn\\
&=&\int d^{2p+2}x\ \bigg(\frac{1}{2e^2(p+1)!}F_{\mu_1\mu_2\cdots\mu_{p+1}}F^{\mu_1\mu_2\cdots\mu_{p+1}}
\nn\\
&&+\frac{\theta}{8\big((p+1)!\big)^2\pi^2}\epsilon^{\mu_1\mu_2\cdots\mu_{p+1}\nu_1\nu_2\cdots\nu_{p+1}}
F_{\mu_1\mu_2\cdots\mu_{p+1}}F_{\nu_1\nu_2\cdots\nu_{p+1}}
\bigg).
\eea
We remind that the theta term only survives when $p$ is an odd number. To consider the electric-magnetic duality with the non-vanishing theta term, we let $p$ be an odd number. 

Now we perform the electric-magnetic duality by introducing the auxiliary field $G$:
\bea
S_p&\rightarrow&\int d^{2p+2}x\ \bigg\lbrack\frac{1}{2e^2(p+1)!}F_{\mu_1\mu_2\cdots\mu_{p+1}}F^{\mu_1\mu_2\cdots\mu_{p+1}}
\nn\\
&&+\frac{\theta}{8\big((p+1)!\big)^2\pi^2}\epsilon^{\mu_1\mu_2\cdots\mu_{p+1}\nu_1\nu_2\cdots\nu_{p+1}}
F_{\mu_1\mu_2\cdots\mu_{p+1}}F_{\nu_1\nu_2\cdots\nu_{p+1}}
\nn\\
&&+\frac{1}{2\big((p+1)!\big)^2\pi}\epsilon^{\mu_1\mu_2\cdots\mu_{p+1}\nu_1\nu_2\cdots\nu_{p+1}}G_{\mu_1\mu_2\cdots\mu_{p+1}}
\nn\\
&&\times\bigg(F_{\nu_1\nu_2\cdots\nu_{p+1}}-\big(\partial_{\nu_1}A_{\nu_2\nu_3\cdots\nu_{p+1}}+
\partial_{\nu_2}A_{\nu_3\nu_4\cdots\nu_{p+1}\nu_1}+\cdots\big)\bigg)\bigg\rbrack,
\eea
in which $F$ is an arbitrary ($p+1$)-form without any constraints and
\bea
\partial_{\nu_1}A_{\nu_2\nu_3\cdots\nu_{p+1}}+
\partial_{\nu_2}A_{\nu_3\nu_4\cdots\nu_{p+1}\nu_1}+\cdots\rightarrow \partial_{\nu_1}A_{\nu_2}-\partial_{\nu_2}A_{\nu_1}
\eea
when $p=1$. We can integrate $F$ out to obtain
\bea
&&\frac{1}{e^2(p+1)!}F_{\mu_1\mu_2\cdots\mu_{p+1}}+\frac{\theta}{4\big((p+1)!\big)^2\pi^2}\epsilon_{\mu_1\mu_2\cdots\mu_{p+1}\nu_1\nu_2\cdots\nu_{p+1}}F^{\nu_1\nu_2\cdots\nu_{p+1}}
\nn\\
&&+\frac{1}{2\big((p+1)!\big)^2\pi}\epsilon_{\mu_1\mu_2\cdots\mu_{p+1}\nu_1\nu_2\cdots\nu_{p+1}}G^{\nu_1\nu_2\cdots\nu_{p+1}}=0
\eea
or
\bea
-\frac{\theta}{4\pi^2}F_{\mu_1\mu_2\cdots\mu_{p+1}}+\frac{1}{e^2(p+1)!}\epsilon_{\mu_1\mu_2\cdots\mu_{p+1}\nu_1\nu_2\cdots\nu_{p+1}}F^{\nu_1\nu_2\cdots\nu_{p+1}}-\frac{1}{2\pi}G_{\mu_1\mu_2\cdots\mu_{p+1}}=0.
\eea
Hence, we could get
\bea
&&\bigg(\frac{1}{e^4(p+1)!}+\frac{\theta^2}{16(p+1)!\pi^4}\bigg)F_{\mu_1\mu_2\cdots\mu_{p+1}}+\bigg(\frac{\theta}{8(p+1)!\pi^3}G_{\mu_1\mu_2\cdots\mu_{p+1}}
\nn\\
&&+\frac{1}{2e^2\big((p+1)!\big)^2\pi}\epsilon_{\nu_1\mu_2\cdots\mu_{p+1}\nu_1\nu_2\cdots\nu_{p+1}}G^{\nu_1\nu_2\cdots\nu_{p+1}}\bigg)=0
\eea
or
\bea
F_{\mu_1\mu_2\cdots\mu_{p+1}}&=&-\frac{\pi}{e^2\bigg(\frac{2(p+1)!\pi^2}{e^4}+\frac{\theta^2(p+1)!}{8\pi^2}\bigg)}\epsilon_{\mu_1\mu_2\cdots\mu_{p+1}\nu_1\nu_2\cdots\nu_{p+1}}G^{\nu_1\nu_2\cdots\nu_{p+1}}
\nn\\
&&-\frac{\frac{\theta}{2\pi}}{\frac{\theta^2}{4\pi^2}+\frac{4\pi^2}{e^4}}G_{\mu_1\mu_2\cdots\mu_{p+1}}.
\eea
After we perform the electric-magnetic duality, we could obtain 
\bea
&&\int d^{2p+2}x\ \bigg(\frac{1}{2e^2(p+1)!}G_{\mu_1\mu_2\cdots\mu_{p+1}}G^{\mu_1\mu_2\cdots\mu_{p+1}}
\nn\\
&&-\frac{\theta}{8\big((p+1)!\big)^2\pi^2}\epsilon^{\mu_1\mu_2\cdots\mu_{p+1}\nu_1\nu_@\cdots\nu_{p+1}}G_{\mu_1\mu_2\cdots\mu_{p+1}}
G_{\nu_1\nu_2\cdots\nu_{p+1}}\bigg)\frac{1}{\frac{\theta^2}{4\pi^2}+\frac{4\pi^2}{e^4}}.
\eea
Thus, the electric-magnetic duality transformation could show:
\bea
\tau\rightarrow\frac{\frac{\theta}{2\pi}}{\frac{\theta^2}{4\pi^2}+\frac{4\pi^2}{e^4}}+\frac{\frac{2\pi i}{e^2}}{\frac{\theta^2}{4\pi^2}+\frac{4\pi^2}{e^4}}=\frac{1}{\frac{\theta}{2\pi}-\frac{2\pi i}{e^2}}=-\frac{1}{\tau}.
\eea
When $p$ is an even number, we only need to set $\theta=0$ and the result is not changed. We define two transformations. The first transformation is the $S$-transformation:
\bea
S(\tau)=-\frac{1}{\tau}
\eea
and the second transformation is the $T$-transformation:
\bea
T(\tau)=\tau+1,
\eea
which generates the transformation:
\bea
\theta\rightarrow\theta+2\pi.
\eea
The combination of the $T$-transformation and the $S$-transformation is an $SL(2)$ transformation.

Now we consider a manifold with a boundary $M$. The action is 
\bea
S_{pb}&=&\frac{1}{2p!\pi}\int_Md^{2p+1}x\ J_{\mu_1\mu_2\cdots\mu_p}B^{\mu_1\mu_2\dots\mu_p}
\nn\\
&&+\frac{i}{8\pi}\int d^{2p+2}x\ \bigg(\tau^*F^+_{\mu_1\mu_2\cdots\mu_{p+1}}F^{+,\ \mu_1\mu_2\cdots\mu_{p+1}}
-\tau F^-_{\mu_1\mu_2\cdots\mu_{p+1}}F^{-,\ \mu_1\mu_2\cdots\mu_{p+1}}\bigg),
\nn\\
\eea
in which $B$ is a $p$-form field lives on the boundary $M$ and is also a boundary value of $A$, which is a gauge potential of $F$ and $J$ is a conserved current on $M$. Because an equation of the motion of $B$ on the boundary $M$ depends on the gauge potential $B$ rather than a field strength, we only perform the electric-magnetic duality on the bulk space with a fixed boundary value of $A$, which is not changed by the electric-magnetic duality \cite{Witten:2003ya}.

Thus, we add 
\bea
&&\frac{1}{2\big((p+1)!\big)^2\pi}\int d^{2p+2}x\
\epsilon^{\mu_1\mu_2\cdots\mu_{p+1}\nu_1\nu_2\cdots\nu_{p+1}}G_{\mu_1\mu_2\cdots\mu_{p+1}}
\nn\\
&&\times\bigg(F_{\nu_1\nu_2\cdots\nu_{p+1}}-\big(\partial_{\nu_1}A_{\nu_2\nu_3\cdots\nu_{p+1}}+
\partial_{\nu_2}A_{\nu_3\nu_4\cdots\nu_{p+1}\nu_1}+\cdots\big)\bigg)\bigg\rbrack
\eea
and $F$ becomes an arbitrary ($p+1$)-form without any restrictions and $G$ is also an arbitrary ($p+1$)-form. We could first integrate $F$ out in the bulk region and obtain
\bea
&&\frac{i}{8\pi}\int d^{2p+2}x\ \bigg(\tau^{\prime *}G^+_{\mu_1\mu_2\cdots\mu_{p+1}}G^{+,\ \mu_1\mu_2\cdots\mu_{p+1}}
-\tau^{\prime} G^-_{\mu_1\mu_2\cdots\mu_{p+1}}G^{-,\ \mu_1\mu_2\cdots\mu_{p+1}}\bigg)
\nn\\
&&-\frac{1}{2\big((p+1)!\big)^2\pi}\int d^{2p+2}x\
\epsilon^{\mu_1\mu_2\cdots\mu_{p+1}\nu_1\nu_2\cdots\nu_{p+1}}G_{\mu_1\mu_2\cdots\mu_{p+1}}
\nn\\
&&\times 
\big(\partial_{\nu_1}A_{\nu_2\nu_3\cdots\nu_{p+1}}+
\partial_{\nu_2}A_{\nu_3\nu_4\cdots\nu_{p+1}\nu_1}+\cdots\big)
\nn\\
&&+\frac{1}{2p!\pi}\int_Md^{2p+1}x\ J_{\mu_1\mu_2\cdots\mu_p}B^{\mu_1\mu_2\dots\mu_p},
\eea
where
\bea 
\tau^{\prime}\equiv-\frac{1}{\tau}.
\eea 
Now $G$ is not an arbitrary $(p+1)$-form after we perform the electric-magnetic duality. Thus, we obtain a generalized BF term:
\bea
&&-\frac{1}{2\big((p+1)!\big)^2\pi}\int d^{2p+2}x\
\epsilon^{\mu_1\mu_2\cdots\mu_{p+1}\nu_1\nu_2\cdots\nu_{p+1}}G_{\mu_1\mu_2\cdots\mu_{p+1}}
\nn\\
&&\times 
\big(\partial_{\nu_1}A_{\nu_2\nu_3\cdots\nu_{p+1}}+
\partial_{\nu_2}A_{\nu_3\nu_4\cdots\nu_{p+1}\nu_1}+\cdots\big)
\nn\\
&\rightarrow&-\frac{1}{2(p!)^2\pi}\int_M d^{2p+1}x\ \epsilon^{\mu_1\mu_2\cdots\mu_{2p+1}} B_{\mu_1\mu_2\cdots\mu_p}\partial_{\mu_{p+1}}A_{\mu_{p+2}\mu_{p+3}\cdots\mu_{2p+1}}.
\eea
In other words, the S-transformation shows
\bea
\tau\rightarrow -\frac{1}{\tau}, \qquad F\rightarrow G,
\eea
and a generalized BF term on the boundary $M$.

The T-transformation ($\theta\rightarrow\theta+2\pi$) could give the generalized form of the Chern-Simons theory:
\bea
\frac{1}{4(p!)^2\pi}\int_Md^{2p+1}x\ \epsilon^{\mu_1\mu_2\cdots\mu_{2p+1}}B_{\mu_1\mu_2\cdots\mu_{p}}\partial_{\mu_{p+1}}
B_{\mu_{p+2}\mu_{p+3}\cdots\mu_{2p+1}}
\eea
from the theta term. This means that the T-transformation is not a symmetry of the bulk. Thus, the boundary theory under the T-transformation should give the generalized form of the Chern-Simons theory:
\bea
-\frac{1}{4(p!)^2\pi}\int_Md^{2p+1}x\ \epsilon^{\mu_1\mu_2\cdots\mu_{2p+1}}B_{\mu_1\mu_2\cdots\mu_{p}}\partial_{\mu_{p+1}}
B_{\mu_{p+2}\mu_{p+3}\cdots\mu_{2p+1}}
\eea
to compensate the non-symmetric part of the bulk theory when $p$ is an odd number. 

If we assume that a duality web on the boundary could be generated from the T-transformation and the S-transformation, we could have the generalized form of the Chern-Simons theory and the generalized BF term when $p$ is an odd number. If $p$ is an even number, we only have the generalized BF term. 

\subsection{$p$ is an Even Number}
Because we only have the generalized form of the Chern-Simons theory, it is hard to use the symmetry breaking to study the boson theory of a duality web and obtain the generalized form of the Chern-Simons theory. Thus, we only study the fermion theory. To give a concrete study, we demonstrate the result of $p=0$. The fermion theory is written as
\bea
&&\int D\psi\ \exp\bigg\lbrack -\int_0^{\beta} dt\ \bigg(\psi^{\dagger}(i\partial_{0}-i\phi+m_f)\psi\bigg)\bigg\rbrack\cdot\exp\bigg(-\frac{i}{2}\int_0^{\beta} dt\ \phi\bigg).
\eea
From the $SL(2)$ transformations in the bulk, we only have the Chern-Simons term. Thus, we could have a topological term, in which the gauge field is a background field. From the combination of the bulk theory and boundary theory at the IR limit, it is easy to know that a background gauge field on the boundary should come from the bulk theory and a dynamical gauge field on the boundary does not. A background gauge field on the boundary needs to couple a fermion field with the parity anomaly and a dynamical gauge fields needs to couple to boson fields. From the demonstration of $p=0$, we possibly lose the boson theory when $p$ is an even number. 

\subsection{$p$ is an Odd Number}
Because we have the generalized form of the Chern-Simons theory and the generalized BF term, it is possible to have a boson theory and a fermion theory simultaneously. The successful example is $p=1$ or the three dimensional duality web. 

Now we consider $p=3$. For this case, we need to consider a seven dimensional boundary theory and an eight dimensional bulk theory. In this case, it is impossible to find a local interacting term that a scalar field coupled to a dynamical gauge field through a naive dimensional analysis. This means that a duality web is hard to generate in higher dimensions.

\section{Conclusion and Outlook}
\label{6}
We discussed a duality web in the condensed matter systems and started from the conjecture \cite{Seiberg:2016gmd} to derive various dualities in three dimensions without losing the holonomy. Our result showed that the dualities are not modified from the non-trivial holonomy. 

We also considered the mean-field analysis with the order parameter $\bar{\psi}\psi$ \cite{Ma:2016yas} in the fermion theory and we found the dual boson theory, which becomes massive. Our result showed an interesting operator correspondence between the boson theory and the fermion theory. The mean-field study is also a powerful tool to study the dynamics. Thus, it should be useful by combining a duality web and the mean-field analysis to study the strongly coupled field theories.

To discuss a phase transition or a phase structure with respect to a temperature, we introduced a temperature in the duality web \cite{Ma:2016yas}. Indeed, the difficulty of introducing a temperature is that the gauge invariant Dirac fermion theory needs to do resummation to all orders, but the duality web may be only guaranteed at  the IR limit or the leading order of the resummation. Thus, we proposed that the duality web at a finite temperature only keeps the Abelian Chern-Simons term with a temperature dependent coefficient. We also found that the fermion theory and the boson theory vanishes when we consider a very high temperature. Thus, the duality web at a finite temperature possibly exists.

We are also interested in extending the duality web to other odd dimensions. To do its extension, we combined the bulk theory and the boundary theory. In three dimensions, the $T$-transformation and the $S$-transformation could generate the duality web \cite{Seiberg:2016gmd}. Thus, we used this method to do our starting point to understand the results of other odd dimensions. We considered the electric-magnetic duality of the $p$-form theory in $2p+2$ dimensions and showed the generalized form of the Chern-Simons theory and the generalized BF term. Our result showed that we possibly lose the boson theory when $p$ is an even number. Only when $p$ is an odd number, it is possible to have a duality web. To find a duality web in higher dimensions, we possibly need to consider a non-local theory. 

Although we need to use a $Spin_c$ manifold to study the three dimensional duality web with a most generic background, it does not mean that we could not have a duality web for some backgrounds. When we studied the three dimensional duality web at a finite temperature and the higher dimensional duality web, in which dimensions are larger than three and are odd, we do not consider the most generic background. Even if a higher dimensional orientable manifold may not admit a $Spin_c$ manifold, this does not mean that a higher dimensional duality web could not exist for some backgrounds. 

One interesting application of our study is to use the mean-field analysis to study the three dimensional duality web at a finite temperature. The powerful part of the duality web is that we could use different ways to study a phase transition and the mean-field method is that we could study dynamics easily. The phase transition with respect to a finite temperature should be interesting and also gives us more understanding to the quantum Hall effect.

The entanglement entropy in a gauge theory is hard to have a tensor product decomposition of a Hilbert space so it is hard to define or study the entanglement entropy. From the three dimensional duality web, we found that the interacting scalar field theory coupled to a dynamical gauge field could dual to the interacting scalar field theory at the IR limit. This means that we could define the entanglement entropy in a theory containing a dynamical gauge field. Although the three dimensional duality web is only guaranteed at an IR limit, a renormalization group flow could help us to extend the definition of the entanglement entropy to other energy scales. Thus, the duality web also helps us to find a suitable definition of the entanglement entropy in a gauge theory.

\section*{Acknowledgments}
The author would like to thank Chang-Tse Hsieh for his useful discussion and Nan-Peng Ma for his encouragement.

\end{document}